\documentclass[11pt]{article}
\usepackage{amsmath,enumerate,fancyhdr,amssymb, amsthm, pstricks, epsf, lscape}

\newcommand{\commentout}[1]{}
\newcommand{\co}[1]{}

\newtheorem{Claim}{Claim}

\co{
\setlength{\oddsidemargin}{0cm} 
\setlength{\evensidemargin}{0cm}
\setlength{\textwidth}{16.0cm} 
\setlength{\topmargin}{1cm}
\setlength{\textheight}{19cm}
}

\def\eref#1{(\ref{#1})}

\newcommand{\norm}[1]{\parallel \! #1 \! \parallel}

\newcommand{\rad}[1]{\mathbb{R}^{#1}}

\newcommand{\zad}[1]{\mathbb{Z}^{#1}}
\newcommand{\qad}[1]{\mathbb{Q}^{#1}}

\newcommand{\nin}{\noindent}

\newcommand{\bx}{\bar{x}}

\newcommand{\lin}{\operatorname{lin}}

\newcommand{\di}{\displaystyle}

\newcommand{\Proj}{\operatorname{Proj}}

\newtheorem{Definition}{Definition}

\newtheorem{Setup}{Setup}
\newtheorem{Example}{Example}
\newtheorem{Counterexample}{Counterexample}
\newtheorem{Proposition}{Proposition}
\newtheorem{Lemma}{Lemma}
\newtheorem{Theorem}{Theorem}
\newtheorem{Corollary}[Definition]{Corollary}
\newtheorem{Remark}[Definition]{Remark}
\newtheorem{Assumption}{Assumption}
\newtheorem{Recipe}{Recipe}

\newcommand{\beq}{\begin{equation}}
\newcommand{\eeq}{\end{equation}}
\newcommand{\beqa}{\begin{eqnarray}}
\newcommand{\eeqa}{\end{eqnarray}}
\newcommand{\ba}{\begin{array}}
\newcommand{\ea}{\end{array}}
\newcommand{\bac}{\begin{array}{ccccccccccc}}
\newcommand{\eac}{\end{array}}
\newcommand{\bprop}{\begin{Proposition}}
\newcommand{\eprop}{\end{Proposition}}

\newcommand{\bcex}{\begin{Counterexample}}
\newcommand{\ecex}{\end{Counterexample}}

\newcommand{\beqast}{\begin{eqnarray*}}
\newcommand{\eeqast}{\end{eqnarray*}}
\newcommand{\benum}{\begin{enumerate}}
\newcommand{\eenum}{\end{enumerate}}
\newcommand{\bit}{\begin{itemize}}
\newcommand{\eit}{\end{itemize}}
\newcommand{\bth}{\begin{Theorem}}
\newcommand{\enth}{\end{Theorem}}

\newcommand{\bdef}{\begin{Definition}}
\newcommand{\Edef}{\end{Definition}}

\newcommand{\bsetup}{\begin{Setup}}
\newcommand{\esetup}{\end{Setup}}
\newcommand{\ble}{\begin{Lemma}}
\newcommand{\ele}{\end{Lemma}}
\newcommand{\bex}{\begin{Example}}
\newcommand{\eex}{\end{Example}}
\newcommand{\bcor}{\begin{Corollary}}
\newcommand{\ecor}{\end{Corollary}}
\newcommand{\brem}{\begin{Remark}}
\newcommand{\erem}{\end{Remark}}
\newcommand{\bass}{\begin{Assumption}}
\newcommand{\eass}{\end{Assumption}}

\newcommand{\brep}{\begin{Recipe}}
\newcommand{\erep}{\end{Recipe}}

\newcommand{\pf}[1]{\vspace{.35cm} \nin {\bf Proof {#1} }}

\newcommand{\bpx}{\begin{pmatrix}}
\newcommand{\epx}{\end{pmatrix}}

\newcommand{\bbx}{\begin{bmatrix}}
\newcommand{\ebx}{\end{bmatrix}}

\begin{document}

% Bala phone: 509-335-3136, 509-334-5450
\title{\bf Branching proofs of infeasibility in low density subset sum problems} 
\author{G\'{a}bor Pataki \thanks{Department of Statistics and Operations Research, UNC Chapel Hill, {\bf gabor@unc.edu}} \ and  
Mustafa Tural \thanks{Department of Statistics and Operations Research, UNC Chapel Hill, {\bf tural@email.unc.edu}} \\
Technical Report 2008-03 \\ Department of Statistics and Operations Research, UNC Chapel Hill}
\date{}

\maketitle

%\tableofcontents

\begin{abstract}

We prove that the subset sum problem 
\beq \label{ss} \tag{$SUB$}
\ba{rcl}
ax & = & \beta \\
x  & \in & \{ 0, 1 \}^n 
\ea
\eeq
has a polynomial time computable certificate of infeasibility for all $a$ with density at most $1/(2n), \,$ 
and for almost all $\beta$ integer right hand sides. 
The certificate is branching on a hyperplane, i.e. by a methodology dual to the one 
explored by Lagarias and Odlyzko \cite{LO85}; Frieze \cite{F86}; Furst and Kannan \cite{FK89}; and Coster et. al. in \cite{CJLOSS92}.

The proof has two ingredients. We first prove that a vector that is near parallel to $a$ is a suitable branching direction, regardless of the 
density. Then we show that for a low density $a$ 
such a near parallel vector can be computed using diophantine approximation, via a methodology introduced by Frank and Tardos in \cite{FT87}. 

We also show that there is a small number of 
long intervals whose disjoint union covers  the integer right hand sides, for which the infeasibility of \eref{ss} is proven by branching on 
the above hyperplane. 

\vspace{0.5cm}

\end{abstract}

\section{Introduction, and main results}
\label{section-intro}
The subset sum problem \eref{ss} is one of the original NP-complete problems introduced by Karp \cite{K72}. A particular reason 
for its importance is its applicability in cryptography. With $a$ being a public key, and $x$ the message, 
one can transmit $\beta = ax \,$ instead of $x$. 
An eavesdropper would need to find $x$ from the intercepted $\beta, \,$ and the public $a$, i.e. solve \eref{ss}, while a legitimate receiver can use a suitable private key 
to decode the message. In cryptography applications, instances with low density are of interest, with 
the density of $a \in \zad{n} \,$ defined as
\beq
d(a) \, = \, \dfrac{n}{\log_2 \norm{a}_\infty}.
\eeq

A line of research started in the seminal paper of Lagarias and Odlyzko  \cite{LO85}, focused on solving such instances. 
%and we also define the density of \eref{ss} as the density of $a$. 
In \cite{LO85} the authors proved that 
the solution \eref{ss} can be found for all but at most a fraction of $1/2^n$ $a$ vectors with $d(a)<c/n, \,$ 
and assuming that the solution exists. Here $c$ is a constant approximately equal to $4.8.$ 
%a solution, and have density less than $c/n$ can be found in polynomial time, where $c \approx 4.8.$ 
%Clearly $d(a) < c/n \,$ is equivalent to $2^{n^2/c} < \norm{a}_\infty$. 
Frieze in \cite{F86} gave a simplified algorithm to prove their result.

From now on we will say that a statement is true for almost all 
elements of a set $S$, if it is true for all, but at most a fraction of $1/2^n$ of them, with the value of $n$ always clear from the context. 

Furst and Kannan in \cite{FK89} pursued an approach that looked at both feasible, and infeasible instances. 
In \cite{FK89} they showed that for some $c > 0$ constant, if $M \geq 2^{c n \log n}, \,$ then for almost 
all $a \in \{\, 1, \dots, M \, \}^n$ and all $\beta$ the problem \eref{ss} has a polynomial size proof of feasibility or infeasibility.
Their second result shows that for some $d > 0$ constant, if $M \geq 2^{d n^2}, \,$ then for almost 
all $a \in \{\, 1, \dots, M \, \}^n$ and all $\beta$ the problem \eref{ss} can be {\em solved} in polynomial time. 

All the above proofs construct a candidate solution to \eref{ss} as a short vector in a certain lattice. Finding a vector
whose length is off by a factor of at most $2^{(n-1)/2}$ from the shortest one is done utilizing the famed basis reduction method of Lenstra, Lenstra, and Lov\'asz \cite{LLL82}. 

Assuming the availability of a {\em lattice oracle}, which finds the shortest vector in a lattice, 
Lagarias and Odlyzko in \cite{LO85} show a similar result under weaker assumption $d(a) < 0.6463.$ The current best result on finding the solution of 
almost all solvable subset sum problems using a lattice oracle is by Coster et al \cite{CJLOSS92}:  they require only $d(a)<0.9408.$ 
It is an open question to prove the infeasibility of almost all subset sum problems with density upper bounded by a constant, 
without assuming the availibility of an oracle. 
For more references, we refer to \cite{CJLOSS92} and \cite{MI02}. 

In this work we look at the structure of low density subset sum problems from a complementary, or dual viewpoint. 
%, which will allow us to further narrow the range of hard instances. 
With $P \,$  a polyhedron and $v$ an integral vector, it is clear that 
$P$ has no integral point, if $vx$ is nonintegral for all $x \in P$. We will examine such proofs of infeasibility of \eref{ss}.
Let
\beqa
G(a,v) & = & \{ \, \beta \in \zad{} \, | \,  vx \not \in \zad{} \; \text{for all} \,\, x \, \text{with} \, ax = \beta, \, 0 \leq x \leq e \, \},
\eeqa
\co{
\beq
\ba{rcl}
G(a,v) & = & \{ \, \beta \in \zad{} \, | \,  vx \not \in \zad{} \; \text{for all} \,\, x \, \text{with} \, ax = \beta, \, 0 \leq x \leq e \, \} \\
       & = & \{ \, \beta \in \zad{} \, | \,  \ell < vx < \ell+1 \; \text{for all} \,\, x \, \text{with} \, ax = \beta, \, 0 \leq x \leq e,  \, \\
       &   &  \;\;\;\;\;\;\;\;\;\;\;\;\;\;\; \text{and for some integer} \; \ell \; \} 
\ea
\eeq
}
where $e$ denotes a column vector of all ones. We will say that for the right hand sides $\beta$ in $G(a,v)$ the infeasibility of \eref{ss} is proven by branching on $vx. \,$ 
The reason for this terminology is that 
letting $P = \, \{ \, x \, | \, ax = \beta, \, 0 \leq x \leq e \, \}, \,$ $\beta$ is in $G(a,v)$ iff the maximum and the minimum of $vx$ over $P$ is between two consecutive integers.
%So given $v, \,$ one needs to solve two linear programs to decide whether the infeasibility of \eref{ss} is proven if $\beta \in G(a,v)$. 

We shall write $\zad{n}_{+},$ and $\zad{n}_{++},$ for the set of nonnegative, and positive integral $n$-vectors, respectively. We will throughout assume $n \geq 10$,
and that the components of $a$ are relatively prime. We only consider nontrivial right hand sides of \eref{ss}, i.e. right hand sides from $\{ \, 0, 1, \dots, \norm{a}_1 \, \}$.
\newpage
\nin Our first main result is:
\bth \label{ss-thm}
Suppose $d(a) \leq 1/(2n)$. Then we can compute in polynomial time an integral vector $v,$ such that 
for almost all right hand sides the infeasibility of \eref{ss} is proven by branching on $vx$. 

Also, $G(a,v)$ can be covered by the disjoint union of at most $2^{2n^2}$ intervals, each of length at least 
$2^{n}$.
\enth
\qed

Note that Theorem \ref{ss-thm} further narrows the range of hard instances from the work of Furst and Kannan in \cite{FK89}. 

There are at most $2^n$ right hand sides for which \eref{ss} is feasible, so most right hand sides lead to an infeasible instance, when 
$d(a)$ is small. 
However, in principle, it may be difficult to {\em prove} the infeasibility of many infeasible instances. Fortunately, this is not the case, as shown by the following corollary.
\bcor \label{cor1}
Let $a$ and $v$ be as in Theorem \ref{ss-thm}. Then for almost all right hand sides for which \eref{ss} is infeasible, its infeasibility is proven
by branching on $vx$.
\ecor
\co{\pf{} Let $I(a)$ be the set of right hand sides for which \eref{ss} is infeasible. 
Theorem \eref{ss-thm} states
\beq
\dfrac{|G(a,v)|}{\norm{a}_1+1} \, \geq \, 1 - \dfrac{1}{2^n}.
\eeq
Since 
\mbox{$I(a) \subseteq \{ \, 0, \dots, \sum_i a_i +1 \, \}, \,$} 
\ref{ss-thm} implies 
\beq \label{2n-2} 
\dfrac{|G(a,v)|}{I(a)} \, \geq \, 1 - \dfrac{1}{2^n};
\eeq
and since $G(a,v) \subseteq I(a), \,$ \eref{2n-2} means the desired conclusion. }
\qed

There is an interesting duality and parallel between the results on low density subset sum in \cite{LO85, FK89, CJLOSS92} and Theorem \ref{ss-thm}.
The proofs in \cite{LO85, FK89, CJLOSS92} work by constructing a candidate solution, while ours by branching, i.e. by a dual method. At the same time, they all rely 
on basis reduction. In our proof we find $v$ by a method of Frank and Tardos in \cite{FT87}, which uses the simultaneous diophantine approximation method 
of Lenstra, Lenstra, and Lov\'asz \cite{LLL82}, which in turn, also uses basis reduction. 

Theorem \ref{ss-thm} will follow from combining Theorems \ref{ssthm2} and \ref{ssthm3} below. 
Theorem \ref{ssthm2} proves that a ``large'' fraction of righ hand sides in \eref{ss} have their 
infeasibility proven by branching on $vx, \,$ if $v$ is relatively short, and near parallel to $a$. Theorem \ref{ssthm3} will show that 
such a $v$ can be found using diophantine approximation, when $d(a) \leq 1/(2n)$. 

\bth \label{ssthm2}
%Let $a$ be a positive, and $v$ a nonnegative integral vector, and $\lambda$ a real number with $a = \lambda v +r, \,$ and $\norm{r}_1/\lambda < 1$. 

Let $\, v \in \zad{n}_+, \, \lambda \in \rad{}, \, r \in \rad{n}$ with $\lambda \geq 1, \, \norm{r}_1/\lambda < 1$, and 
$$
a = \lambda v +r.
$$
Then the infeasibility of all, but at most a fraction of 
\beq \label{gav} 
\dfrac{2 (\norm{r}_1 +1)}{\lambda}
\eeq
right hand sides is proven by branching on $vx$. 
\co{
\beq \label{gav} 
\dfrac{|G(a,v)|}{\norm{a}_1} \, \geq \, 1 - \dfrac{2 \norm{r}_1 +1}{\lambda}.
\eeq}

In addition, $G(a,v)$ can be covered by the disjoint union of at most $\norm{v}_1$ intervals, each of length at least $\lambda - \norm{r}_1$. 
\enth
\qed

\bth \label{ssthm3}
Suppose $d(a) \leq 1/(2n). \,$ Then we can compute in polynomial time $v \in \zad{n}_+, \, \lambda \in \qad{}, \, r \in \qad{n}$ with $a = \lambda v +r,$ 
and 
\benum
\item \label{1} $\norm{v}_1  \leq  2^{2n^2}$; 
\item \label{2} $\norm{r}_1/\lambda  \leq  1/2^{n+2}$;
\item \label{3} $\lambda \geq  2^{n+2}$.
\eenum
\enth
\qed
%\pf{of Theorem \ref{ss-thm}} 

\brem
{\rm 
In this discussion we clarify what we mean by the $v$ vector of  Theorem \ref{ssthm3} being near parallel to $a$.
%(the fact that it is shorter than $a$ is implied by the lower bound on $\lambda$). 

Given $v, \lambda, \,$ and $r \,$ in Theorem \ref{ssthm3}, assume 
%that $\lambda v$ is the projection of $a$ on the line spanned by $v$. 
%we can assume %that $\lambda v \,$ is the projection of $a$ onto the line spanned by $v$. 
\beq \label{lambdav}
\lambda v \, = \, \Proj \, \{ \, a \, | \, \lin \, \{ v \} \}, \, r = a - \lambda v.
\eeq
\co{If \eref{lambdav} does not hold originally, by making this modification we will
not decrease $\lambda, \,$ and will not increase $\norm{r}. \,$ So we will increase $\norm{r}_1/\lambda \,$ by a factor of at most 
$\sqrt{n},$ and in using $v$ in Theorem \ref{ssthm2} what matters is that $\lambda$ is large, and $\norm{r}_1/\lambda$ small.
%this fact will }
}
Then %Given \eref{lambdav}, we have 
\beq \label{sinav} 
\sin(a,v) \, = \, \dfrac{\norm{r}}{\norm{a}} \, \leq \, \dfrac{\norm{r}}{\norm{\lambda v}} \, \leq \, \dfrac{\norm{r}}{\norm{\lambda}}.
\eeq
So a small upper bound on $\norm{r}/\lambda$ will force $\sin(a,v)$ to be small as well, i.e. $v$ to be near parallel to $a$. 
Some of the inequalities in \eref{sinav} can be strict. For instance, letting $a = (m^2, m^2+1), \, v = (m, m+1), \,$ and defining $\lambda$ and $r$ as 
in \eref{lambdav}, it is easy to check that $r/\lambda \rightarrow (1/2, - 1/2), \, $ as $m \rightarrow \infty, \,$ but obviously $\sin(a,v) \rightarrow 0.$ 
}
\erem

\section{Proofs}

\pf{of Theorem \ref{ssthm2}} Let us fix $a$ and $v$. 
Since $a$ and $v$ are nonnegative, and $e$ is a column vector of all ones, it holds that 
$$
\norm{a}_1 = ae, \, \text{and} \, \norm{v}_1 = ve,
$$
and we will use the latter notation for brevity. 

For a row-vector $w, \,$ and an integer $\ell \,$ we write
\beq
\ba{rcl}
\max(w,\ell) & = & \max \, \{ \, wx \, | \, vx \leq \ell, \, 0 \leq x \leq e \, \}, \\
\min(w,\ell) & = & \min \, \{ \, wx \, | \, vx \geq \ell, \, 0 \leq x \leq e \, \}.
\end{array}
\eeq
The dependence on $v, \,$ and on the sense of the constraint (i.e. 
$\leq, \,$ or $\geq \,$) is not shown  by this notation; however,
we always use $vx \leq \ell \,$ with ``max'', and $vx \geq \ell \,$ with ``min'', and $v$ is fixed. 

\begin{Claim} \label{minmax-claim}
We have 
\beqa \label{minmax1}
\min(a,k)  & \leq  & \max(a,k) \,\, \text{for} \,\, k \in \{0, \dots, ve \}, \,  \\
\label{maxmin}
\max(a, k) - \min(a,k) & \leq & \norm{r}_1 \,\, \text{for} \,\, k \in \{0, \dots, ve \}, \, \text{and} \\
\label{minmax2}
\min(a, k+1) - \max(a,k) & \geq & - \norm{r}_1 + \lambda > 0 \,\, \text{for} \,\, k \in \{0, \dots, ve-1 \}.
\eeqa
\end{Claim}
\pf{} The feasible sets of the optimization problems defining $\min(a,k), \,$ and $\max(a,k) \,$ contain $\{ \, x \, | \, vx=k, \, 0 \leq x \leq e \, \}, \,$ so 
\eref{minmax1} follows. 

\nin The decomposition of $a$ shows that for all $\ell_1$ and $\ell_2$ integers for which the expressions below are defined, 
\beq \label{max-and-min}
\ba{rcl}
\max(a,\ell_1)    & \leq & \max(r,\ell_1) + \lambda \ell_1,  \; \text{and} \\ %\;\; k \in \{0, \dots, \la p, e \ra \} \\
\min(a,\ell_2)    & \geq & \min(r,\ell_2) + \lambda \ell_2,
\end{array}
\eeq
hold. Therefore 
\beq \label{minmax-l12}
\ba{rcl}
\min(a, \ell_2) - \max(a,\ell_1) & \geq & \min(r, \ell_2) - \max(r,\ell_1) + \lambda(\ell_2 - \ell_1) \\
                         & \geq & - \norm{r}_1 + \lambda(\ell_2 - \ell_1).
\end{array}
\eeq
follows, and \eref{minmax-l12} with $\ell_2 = \ell_1 = k \,$ implies \eref{maxmin}, and with $\ell_2 = k+1, \, \ell_1 = k \,$ yields 
\eref{minmax2}. 

\nin Hence
\beq \label{zerope}
\min(a,0)  \leq \max(a,0)  <  \min(a,1) \leq  \max(a,1) < \min(a,2) \leq \dots < \min(a, ve) \leq \max(a, ve).
\eeq
\nin We will call the intervals 
$$
[\min(a, 0), \max(a, 0)], \dots, [\min(a,  ve ), \max(a, ve)]
$$
{\em bad}, 
and the  intervals 
$$
G_0 := (\max(a, 0), \min(a, 1)), \dots, G_{ve-1} := (\max(a, ve -1), \min(a, ve))
$$
{\em good}. 
%\end{document}

The nonnegativity of $v \,$ and of $a \,$ imply
$
\min(a,0) = 0, \, \text{and} \, \max(a, ve) = ae,
$
so the bad, and good intervals partition $[0, ae]$: the pattern is bad, good, \dots, good, bad.
Some of the bad intervals may have zero length, but by \eref{minmax2} none of the good ones do.
%\end{document}

Next we show that the good intervals contain exactly the right hand sides for which the infeasibility of \eref{ss} is proven by branching on $vx$.
\begin{Claim} \label{good-claim} 
\beqa
G(a,v) \, = \, \di{\cup_{i=0}^{ve-1} G_i \cap \zad{}.}
\eeqa
\end{Claim} 
\pf{} By definition $\beta \in G(a,v)$ iff for some $\ell$ integer with $0 \leq \ell < ve-1,$ and for all $x$ with $0 \leq x \leq e, \, ax=\beta \,$ 
\beq
\ell < vx < \ell+1
\eeq
holds. We show that for this $\ell$ 
\beqa \label{maxell} 
\max(a, \ell) & < & \beta \;\; \text{and} \\ \label{minellp1} 
\min(a, \ell+1) & > & \beta.
\eeqa
First, assume to the contrary that \eref{maxell} is false, i.e. there exists $x_1$ with 
\beq \label{x1} 
a x_1 \geq \beta, \, v x_1 \leq \ell, \, 0 \leq x_1 \leq e.
\eeq
Since $\ell \geq 0, \,$ denoting by $x_2$ the all-zero vector, it holds that 
\beq \label{x2} 
a x_2 \leq \beta, \, v x_2 \leq \ell, \, 0 \leq x_2 \leq e.
\eeq
Looking at \eref{x1} and \eref{x2} it is clear that a convex combination of $x_1$ and $x_2, \,$ say $\bx$ satisfies 
\beq \label{bx} 
a \bx = \beta, \, v \bx \leq \ell, \, 0 \leq \bx \leq e,
\eeq
which contradicts \eref{maxell}. Showing \eref{minellp1} is analogous. 

\nin{\bf End of proof of Claim \ref{good-claim}}

To summarize, Claim \ref{good-claim} implies that $G(a,v)$ is covered by the disjoint union of $ve$ intervals. By \eref{minmax2} their length is lower bounded by 
$\lambda - \norm{r}_1.$ 

Let us denote by $b$ the number of integers in bad intervals, and by $g$ the number of 
integers in good intervals, i.e. $g = | G(a,v)|$.
Using \eref{maxmin} and \eref{minmax2}, and  the fact that 
there are $ve \,$ good intervals, and $ve + 1 \,$ bad ones, we get 
\beq
\ba{rcl}
g & \geq & ve (\lambda   - \norm{r}_1 - 1), \\
b & \leq & (ve+1) (\norm{r}_1+1),
\end{array}
\eeq
so 
\beqa
\dfrac{g}{b} & \geq & \dfrac{ve}{ve+1} \dfrac{\lambda   - (\norm{r}_1 + 1)}{\norm{r}_1+1} \\
             & \geq & \dfrac{1}{2} \dfrac{\lambda   - (\norm{r}_1 + 1)}{\norm{r}_1+1} \\
             & \geq & \dfrac{\lambda}{2 (\norm{r}_1+1)} -1,
\eeqa
and from here 
\beqa
\dfrac{b}{g+b} & \leq &  \dfrac{1}{1+ g/b} \\
               & \leq & \dfrac{ 2 (\norm{r}_1+1)}{\lambda}.
\eeqa
follows.
\qed

\pf{of Theorem \ref{ssthm3}} 

We will use a methodology due to Frank and Tardos introduced in \cite{FT87}. 
Here the authors employ simultaneous diophantine approximation to decompose a vector with large norm into the weighted sum of 
smaller norm vectors. We will only need  one vector that approximates $a, \,$ and 
the parameters will be somewhat differently chosen in the diophantine approximation. 

We will rely on the following result of Lenstra, Lenstra, and Lov\'asz from \cite{LLL82}: 

\bth \label{lll}
Given a positive integer $N, \,$ and $\alpha \in \qad{n}, \,$ we can compute in polynomial time $v \in \zad{n}, \, q \in \zad{}_{++} \,$ such that 
\beqa \label{qalpha}
\norm{ q \alpha - v}_\infty & \leq & \dfrac{1}{N} \; \text{and}  \\ 
\label{qN} 
q                           & \leq & 2^{n(n+1)/4} N^n.
\eeqa
\enth
\qed

We will use Theorem \ref{lll} with 
$$
\alpha = \dfrac{a}{\norm{a}_\infty}, \,
$$
then set 
$$
\lambda = \dfrac{\norm{a}_\infty}{q}, \, r = a - \lambda v.
$$
We have the following estimates with ensuing explanation: 
\beqa \label{vnorm}
\norm{v}_1 & \leq & n \norm{v}_\infty \, \leq \, n q \, \leq \,  n 2^{n(n+1)/4} N^n, \\
\label{rnorm} 
\dfrac{\norm{r}_1}{\lambda} & \leq & \dfrac{n \norm{r}_\infty}{\lambda} \, \leq \, \dfrac{n}{N}, \\
%\eeqa
%and 
%\beqa
\label{lambdageq} 
\lambda & \geq & \dfrac{\norm{a}_\infty}{2^{n(n+1)/4} N^n} \, \geq \, \dfrac{2^{2n^2 - n(n+1)/4}}{N^n}. 
\eeqa
Here \eref{vnorm} follows from using \eref{qalpha}, since $\norm{ q \alpha}_\infty = q, \,$ and $v$ is integral. 
The second inequality in \eref{rnorm} is actually equivalent to \eref{qalpha}; and \eref{lambdageq} 
comes from the definition of $\lambda, \,$ and \eref{qN}. 
Hence (1), (2), and (3) in Theorem \ref{ssthm3} are satisfied when 
\beqa \label{N1}
n 2^{n(n+1)/4} N^n  & \leq & 2^{2n^2}, \\ \label{N2} 
\dfrac{n}{N} & \leq & \dfrac{1}{2^{n+2}}, \\ \label{N3} 
\dfrac{2^{2n^2 - n(n+1)/4}}{N^n} & \geq & 2^{n+2}. 
\eeqa
But \eref{N1} through \eref{N3} are equivalent to 
\beq
n 2^{n+2} \, \leq \, N \, \leq \, 2^{2n - (n+1)/4 - 1 - 2/n},
\eeq
and such an integer $N$ exists, when $n \geq 10$. 
\qed

\pf{of Corollary \ref{cor1} } 
Let $I(a)$ be the set of right hand sides for which \eref{ss} is infeasible. 
Theorem \ref{ss-thm} states
\beq
\dfrac{|G(a,v)|}{\norm{a}_1+1} \, \geq \, 1 - \dfrac{1}{2^n}.
\eeq
Since 
\mbox{$I(a) \subseteq \{ \, 0, \dots, \norm{a}_1 \, \}, \,$} 
Theorem \ref{ss-thm} implies 
\beq \label{2n-2} 
\dfrac{|G(a,v)|}{I(a)} \, \geq \, 1 - \dfrac{1}{2^n};
\eeq
and since $G(a,v) \subseteq I(a), \,$ \eref{2n-2} means the desired conclusion. 
\qed

\brem
{\rm 
One can use a different methodology  to find a near parallel vector to $a$, which we quote from \cite{PT07}: 

\bth \label{near-parallel}
\label{pa-1}
Suppose $d(a) \leq 1/(n/2+1)$. 
%$\norm{a} \, \geq \, 2^{(n/2 +1)n}. \,$ 
Let $U$ be a unimodular matrix such that the columns of 
$$
\bpx
a \\
I
\epx U 
$$
are reduced in the sense of Lenstra, Lenstra, and Lov\'asz, and $v$ the last row of $U^{-1}$. 
Define $r$ and $\lambda$ to satisfy \eref{lambdav}, and let 
$f(a) \, = \, 2^{n/4}/\norm{a}^{1/n}.$ 

\nin Then 
\benum
\item \label{pa-1-1} $\norm{v} (1+\norm{r}^{2})^{1/2} \leq \norm{a} f(a)$;
\item \label{pa-1-2} $\lambda \geq 1/f(a)$; %\dfrac{1}{f(a)}$; 
\item \label{pa-1-3} $\norm{r}/\lambda \leq 2 f(a)$.
\eenum
\enth
\qed

These bounds also suffice\co{So this result also yields bounds on $\norm{r}/\lambda \,$ and $\lambda$ which suffice} 
to prove the first part of Theorem \ref{ss-thm}; however, the bound we get on $\norm{v}$ involves $\norm{a}$ as well, not just the dimension.
}
\erem

\nin{\bf Acknowledgement} Thanks are due to Ravi Kannan, and Laci Lov\'asz for helpful discussions; to Fritz Eisenbrand for discussions on the connection 
with diophantine approximation; and to  Jeff Lagarias and Andrew Odlyzko for pointing out reference \cite{CJLOSS92}.

\bibliography{IP_Refs}

\end{document}